\documentclass[acmlarge]{acmart}

\usepackage{booktabs} 
\usepackage{subcaption}
\usepackage{fancyhdr}

\usepackage[ruled]{algorithm2e} 

\SetAlFnt{\small}
\SetAlCapFnt{\small}
\SetAlCapNameFnt{\small}
\SetAlCapHSkip{0pt}
\IncMargin{-\parindent}

\usepackage{color}

\acmDOI{0000001.0000001}

\fancypagestyle{empty}{%
  \fancyhf{}
  \fancyhead[L]{\small {\color{blue} This is a preprint of an article accepted for publication in \textit{Journal of the Association for Information Science and Technology copyright @ 2018 (Association for Information Science and Technology)}}}
}

\begin{document}

\title[A Longitudinal Assessment of the Persistence of Twitter Datasets]{A Longitudinal Assessment of the Persistence of Twitter Datasets}

\author{Arkaitz Zubiaga}
\orcid{0000-0003-4583-3623}
\affiliation{%
  \institution{University of Warwick}
  \streetaddress{Gibbet Hill Road}
  \city{Coventry}
  \postcode{CV4 7AL}
  \country{United Kingdom}}

\begin{abstract}
 With social media datasets being increasingly shared by researchers, it also presents the caveat that those datasets are not always completely replicable. Having to adhere to requirements of platforms like Twitter, researchers cannot release the raw data and instead have to release a list of unique identifiers, which others can then use to recollect the data from the platform themselves. This leads to the problem that subsets of the data may no longer be available, as content can be deleted or user accounts deactivated. To quantify the impact of content deletion in the replicability of datasets in a long term, we perform a longitudinal analysis of the persistence of 30 Twitter datasets, which include over 147 million tweets. Having the original datasets collected between 2012 and 2016, and recollecting them later by using the tweet IDs, we look at four different factors that quantify the extent to which recollected datasets resemble original ones: completeness, representativity, similarity and changingness. Even though the ratio of available tweets keeps decreasing as the dataset gets older, we find that the textual content of the recollected subset is still largely representative of the whole dataset that was originally collected. The representativity of the metadata, however, keeps decreasing over time, both because the dataset shrinks and because certain metadata, such as the users' number of followers, keeps changing. Our study has important implications for researchers sharing and using publicly shared Twitter datasets in their research.
\end{abstract}

\keywords{social media, twitter, datasets, reproducibility}

\maketitle

\renewcommand{\shortauthors}{A. Zubiaga}

\thispagestyle{empty}

\section{Introduction}
\label{sec:introduction}

Archiving and sharing data for future use is a key requirement to enable reproducibility of scientific research \cite{peng2006reproducible,borgman2012conundrum,nosek2015promoting,peng2011reproducible}. In recent years, the use of large-scale datasets for research has increased with the emergence of the Web and social media \cite{ruths2014social,tufekci2014big,kjellberg2016researchers}.
Despite the increasing interest in sharing social media datasets \cite{wallis2013if,stvilia2017toward}, there is also concern about the persistence of these data \cite{weller2016manifesto}. The long-term persistence of datasets can be impeded by different circumstances. One reason is that some datasets become unavailable short after publication due to the lack of stability of the hosting servers \cite{mieskes2017quantitative}; proper use of data repositories such as Figshare.com and Dataverse.org can help mitigate this issue \cite{austin2015research}. Another important reason is that datasets gathered from third party services, such as social media data \cite{chinthakayala2014comparative,hutton2015making,hutton2015towards,metaxas2014sifting}, cannot always be shared in its final form owing to restrictions in the terms and conditions of the data provider. To circumvent these restrictions, researchers can share unique identifiers of individual posts, which then enable recollection of the data, as is the case with Twitter; recollecting social media datasets from unique identifiers, however, leads to the collection of a dataset that differs from the original, as part of the data may no longer be available \cite{mccreadie2012building,weller2016manifesto}, e.g. because user accounts may have been deleted or suspended \cite{liu2014tweets}. Moreover, the metadata of social media posts can also change over time as their authors update their profile, their number of followers or friends keeps changing and/or the number of shares/likes of posts is altered \cite{holsapple2014business}. Consequently, the dataset collected from a publicly shared list of unique identifiers such as tweet IDs will often lead to a subset of the original dataset, with the additional caveat that the metadata of the collected posts may have changed. However, the researcher recollecting the dataset cannot estimate the extent to which the recollected dataset resembles the data compiled by the original authors; this study makes a longitudinal assessment of the degree to which datasets recollected from tweet IDs resemble original datasets.

Previous work has looked into the predictability of tweets being deleted short after being posted, for instance for regretting their content or for having made spelling mistakes \cite{petrovic2013wish,bagdouri2015predicting,potash2016using}. Work has also looked at the probability of a Twitter account being suspended for being offensive or spammer \cite{thomas2011suspended}. While all of these affect the persistence of a Twitter dataset in the long term, no study has looked at the effect of the disappearance of tweets in the persistence of historical Twitter datasets and there is no assessment of the replicability of datasets. Our ultimate goal is to determine how comparable the recollected dataset is. To tackle this, our study sets forth the following four research questions:

\begin{description}
 \item[RQ1] How complete is a dataset recollected from lists of publicly available tweet IDs? (\textit{completeness})
 \item[RQ2] How representative is the recollected dataset? (\textit{representativity})
 \item[RQ3] How similar is the recollected subset to a randomly sampled subset? (\textit{similarity})
 \item[RQ4] How has the metadata associated with the successfully recollected subset changed? (\textit{changingness})
\end{description}

We perform a large-scale, longitudinal analysis of 30 Twitter datasets associated with real world events between 2012 and 2016, which include over 147 million tweets. In May 2016, we simulated the scenario in which a researcher wants to reuse these datasets; we recollected all these datasets using the underlying Twitter IDs, which enabled us to have two versions of all 30 datasets: the dataset originally collected through the streaming API in real-time while the event was happening, and the dataset recollected later on. The study of datasets that had been collected between days earlier and up to four years earlier enables us to analyse the effect of recollecting tweets after different periods of time. This in turn enables us to assess whether and the extent to which the persistence of a Twitter dataset fades. Our analysis shows that the ratio of available tweets keeps decreasing as the dataset gets older; still, interestingly we find that the textual content of the recollected subset is largely representative of the original dataset, irrespective of the dataset's age, which is positive for research that focuses on the textual content of tweets. Results are not as good when we analyse metadata of the recollected datasets; we observe that most of the metadata of the recollected subsets are no longer representative of the original data, in part because some of these metadata have changed over time. However, metadata are found to be to some extent similar to a randomly sampled subset of the original dataset, and therefore it should be carefully used by acknowledging that it is a subset of the original data. These findings present import implications for the practice of sharing and archiving Twitter datasets, as well as for researchers reusing publicly shared Twitter datasets in their research.

\section{Limitations of Sharing Twitter Datasets}
\label{sec:limitations}


With the increasing use of social media data for research purposes \cite{kaplan2010users}, the concerns for making sure that datasets used in these works are released are growing \cite{weller2016manifesto}. Sharing data is crucial for many reasons, including the ability to reproduce results from existing research, as well as to extend existing work with new analyses and findings by using common datasets \cite{borgman2012conundrum}. Results of Twitter experiments can often be largely dependent on the datasets being used. In a study trying to replicate 10 Twitter experiments using newly collected datasets, \citeA{liang2015testing} were unable to replicate the findings of six of them owing mainly to three reasons: lack of detail of the analytical methods utilised, use of inconsistent measurement approaches, and inability to replicate the dataset. Indeed, an ability to reuse existing datasets is key to enable reproducibility and to build on existing work. The sharing of Twitter datasets is however hindered by the need to comply with the platform's Terms of Service\footnote{\url{https://twitter.com/tos}} (TOS). As part of Twitter's TOS, using its API to retrieve data from Twitter requires agreeing with its developer policy\footnote{\url{https://dev.twitter.com/overview/terms/agreement-and-policy}}. The clause F.2 of this policy states that \textit{``If you provide Content to third parties, including downloadable datasets of Content or an API that returns Content, you will only distribute or allow download of Tweet IDs and/or User IDs''}. An exception to that clause is that it is allowed to \textit{``provide export via non-automated means of up to 50,000 public Tweets and/or User Objects per user of your Service, per day''}. While the latter has enabled distribution of small Twitter datasets (e.g. \citeA{zubiaga2016tweetlid}), it is generally not a viable solution to release large-scale datasets, e.g. those associated with major events and breaking news, which clearly exceed the 50,000 tweet limit. Beyond the scientific community, there have been multiple attempts to build permanent archives of deleted tweets, including Undetweetable, Tweleted, Deadbird and PostGhost; however, these have been discontinued following requests from Twitter to take down their service. The only such service that exists today is Politwoops\footnote{\url{https://www.politwoops.eu/}}, which is however limited as it is only allowed to store tweets deleted by politicians.

Given this restriction as per the Twitter's TOS, the workaround that enables researchers to share a large-scale Twitter dataset is hence to share the IDs of all the tweets in the dataset. Having those tweet IDs, another researcher can then go back to Twitter's REST API to retrieve the tweets by using those IDs; the API's \textit{statuses/lookup}\footnote{\url{https://dev.twitter.com/rest/reference/get/statuses/lookup}} method returns the content and metadata of the tweets, with the exception of those that are no longer available. The TREC 2011 Twitter dataset \cite{ounis2011overview} was the first to be released in this way, sharing only the tweet IDs. This methodology was later documented in \citeA{mccreadie2012building}, finding that within a period of 6 months, 27,3\% of the tweets conforming the dataset had become unavailable. Similarly, others have reused existing datasets for their research, for which tweet IDs had been published; they then used those tweet IDs to reconstruct the datasets for their work. For instance, \citeA{hasan2016twitternews} found that around 30\% of the tweets from an old Twitter dataset were unavailable by the time they collected it; they were, however, unable to determine how comparable the remainder 70\% of the dataset was. This issue also affects the organisation of shared tasks, as the organisers release the dataset to be used in the challenge through tweet IDs. In recent shared tasks, \citeA{jaggi2014swiss} found that participants could not retrieve between 10\% and 15\% of the tweets, which in the shared task by \citeA{alegria2015tweetnorm} was slightly lower, 9.4\%. The common practice for making sure that all participants are evaluated against the same set of tweets is for organisers to rely on the subset of tweets available after the evaluation period comes to an end. These datasets are generally publicly released after the shared task finishes and, while it is expected that tweets will continue disappearing, the utility of these datasets as time goes on is uncertain. As a workaround, it has been proposed by \cite{sequiera2017finally} to create annotated datasets by using the `Twitter Stream Grab' datasets available through the Web Archive\footnote{\url{https://archive.org/details/twitterstream}}, where tweets collected through Twitter's Spritzer streaming API\footnote{The Spritzer API provides a 1\% random sample of public tweets.} are archived. While this can be useful in some cases, such as for shared tasks, it is generally insufficient for the study of events through Twitter, as the Spritzer sample does not include all tweets related to those events. In this work we are particularly interested in Twitter datasets associated with events.

There has been research looking into the deletion of social media posts and more specifically tweets \cite{almuhimedi2013tweets}. Some of this work has conducted exploratory research, such as \citeA{bhattacharya2016characterizing,mondal2016forgetting,mondal2017managing} analysing linguistic characteristics of deleted tweets as well as personality traits of authors who delete tweets, and \citeA{almuhimedi2013tweets} analysing the types of tweets that get deleted, looking at different aspects such as the client used for tweeting, sentiment and days of the week. Others tried to predict what tweets would be eventually deleted; prediction of deleted tweets has proven challenging, with performance scores roughly reaching 0.2 in terms of F1 score \cite{petrovic2013wish,bagdouri2015predicting}. \citeA{potash2016using} achieved higher performance scores around 0.45 F1 score, however as the authors acknowledge their dataset had a significantly higher rate of deleted tweets (23\% over 3\% in previous work). Others have analysed suspension \cite{thomas2011suspended}, deletion \cite{volkova2017identifying} and hijacking \cite{thomas2014consequences} of Twitter accounts. Research in this direction can be informative for instance for: (1) understanding and predicting tweets that are likely to be deleted, to inform users in advance that they may regret posting a tweet \cite{sleeper2013read,wang2011regretted,xu2013examination,zhou2015identifying,zhou2016tweet}, (2) identifying content and behaviour that leads to account suspension, so that users can avoid it in the future, and (3) identifying malicious accounts to flag them as such.

Despite the evidence of tweets being deleted, previous work has not quantified the impact of these deleted tweets on existing datasets and hence has not assessed quantitatively the extent to which a reconstructed dataset resembles the original dataset. The present work advances research in this direction by looking at 30 large-scale Twitter datasets collected between 2012 and 2016, comparing datasets collected originally in real-time as well as at a later stage.

\section{Experiment Design}
\label{sec:experiment-design}

Our experiments for assessing the persistence of Twitter datasets consist of three steps: (1) real-time collection of Twitter datasets between 2012 and 2016, (2) recollection of datasets in May 2016, and (3) analysis and quantification of differences.

\subsection{Original Datasets}
\label{ssec:datasets}

Between February 2012 and May 2016, we collected 30 Twitter datasets associated with different real-world events. These include a range of different types of events such as breaking news stories, emergencies, elections and sporting events. The tweets for all these datasets were collected using the same methodology; we employed Twitter's streaming API with a set of relevant keywords and hashtags associated with each event. The use of the streaming API is the most widely used approach for Twitter data collection \cite{zubiaga2017detection}, which enables collection of data in real-time as tweets are posted and hence unaffected by later deletions of tweets.

A summary of these datasets is shown in Table \ref{tab:datasets}, with the time frame and number of tweets in each case. The tweet IDs conforming all these datasets as well as the keywords we used for the collection are publicly available\footnote{\url{https://figshare.com/articles/Twitter_event_datasets_2012-2016_/5100460}}.

\begin{table}
 \centering
 \small
 \begin{tabular}{ l l l l r }
  \toprule
  Event & Year & Start & End & No. tweets \\
  \midrule
  Superbowl & 2012 & February 3 & February 7 & 1,659,475 \\
  SXSW & 2012 & March 8 & March 22 & 1,563,448 \\
  Euro 2012 & 2012 & June 2 & July 4 & 8,992,157 \\
  Mexican election & 2012 & July 1 & July 3 & 191,788 \\
  Hurricane Sandy & 2012 & October 25 & November 5 & 14,914,566 \\
  Obama \& Romney & 2012 & November 5 & November 8 & 10,146,517 \\
  US election & 2012 & November 5 & November 8 & 1,740,258 \\
  
  Boston Marathon bombing & 2013 & April 15 & April 16 & 3,430,387 \\
  
  St. Patrick's Day & 2014 & March 15 & March 18 & 2,882,010 \\
  Gaza under attack & 2014 & June 1 & July 18 & 2,886,322 \\
  Ebola outbreak & 2014 & July 1 & July 31 & 986,525 \\
  Ferguson unrest & 2014 & August 9 & August 26 & 8,782,071 \\
  Indyref & 2014 & September 17 & September 20 & 1,524,166 \\
  Hong Kong protests & 2014 & September 26 & October 20 & 1,188,372 \\
  Ottawa shooting & 2014 & October 22 & October 24 & 1,075,864 \\
  Typhoon Hagupit & 2014 & December 5 & December 11 & 264,626 \\
  Sydney siege & 2014 & December 14 & December 17 & 2,157,879 \\
  
  Charlie Hebdo shooting & 2015 & January 7 & January 14 & 1,894,0619 \\
  Germanwings plane crash & 2015 & March 24 & March 30 & 2,648,983 \\
  Nepal Earthquake & 2015 & April 25 & May 18 & 12,004,187 \\
  Refugees Welcome & 2015 & September 2 & November 24 & 1,743,153 \\
  Hurricane Patricia & 2015 & October 24 & December 8 & 1,151,220 \\
  Paris Attacks & 2015 & November 13 & November 24 & 29,821,274 \\
  
  Irish election & 2016 & February 3 & March 6 & 758,803 \\
  Brexit & 2016 & February 24 & May 3 & 1,826,290 \\
  Brussels Airport explossion & 2016 & March 22 & March 30 & 5,869,990 \\
  Lahore blast & 2016 & March 27 & March 30 & 1,149,253 \\
  Cyprus hijacked plane & 2016 & March 29 & March 30 & 702,586 \\
  Panama papers & 2016 & April 3 & May 3 & 5,044,379 \\
  Sismo Ecuador & 2016 & April 17 & April 28 & 1,007,867 \\
  \midrule
  \textbf{TOTAL} &  &  &  & \textbf{147,055,035} \\
  \bottomrule
 \end{tabular}
 \caption{Time frames and number of tweets in the datasets under study.}
 \label{tab:datasets}
\end{table}

\subsection{Recollection of Datasets}
\label{ssec:recollection}

The next step consisted in simulating the scenario of the researcher who would be using the tweet IDs to recollect the datasets. In May 7th, 2016, we used Twitter's API to recollect the 30 aforementioned datasets. Having the over 147 million tweet IDs as input, we used the API's \textit{statuses/lookup} method\footnote{\url{https://dev.twitter.com/rest/reference/get/statuses/lookup}} to retrieve tweets by ID. To make sure that our recollection methodology did not benefit any of the datasets, we collected all of them in parallel. The parallelisation of the tweet recollection enabled us to complete the task within three weeks by May 28th. At this point, we have two versions of each dataset: $D_o$, the dataset originally collected while the event was occurring, and $D_r$, the dataset recollected in May 2016.

\subsection{Assessing the Persistence of Datasets}
\label{ssec:comparison}

With the two versions we collected for each of the 30 datasets, $D_{o\{1..30\}}$ and $D_{r\{1..30\}}$, we use four different metrics to measure the extent to which recollected datasets retain their original characteristics. These four metrics enable us to measure the completeness, representativity, similarity and changingness of each of the recollected datasets $D_{ri}$ with respect to the original $D_{oi}$, where $i \in [1, 30]$. In the next section we describe the approach we followed to measure each of the four characteristics and present the analysis of results.

\section{Analysis}
\label{sec:analysis}

This section is organised by research question in four parts: completeness, representativity, similarity and changingness.

\subsection{Completeness}
\label{completeness}

With the completeness we measure the extent to which the recollected datasets resemble the original ones in size. That is, how many of the elements in the original dataset can still be found in the recollected dataset. We look at six different types of elements: (1) tweets, (2) unique hashtags, (3) unique user mentions, (4) unique tokens found in the tweets' contents, (5) unique URLs and (6) unique tweet authors.

A first look at the total number of successfully recollected tweets shows that 119,752,714 tweets (81.4\% of the whole) were still available. These percentages are more favourable for the majority of the elements, with 86.8\% of the hashtags available, 84.5\% of the user mentions, 86.5\% of the tokens and 89.8\% of the URLs. The exception is the percentage of unique users found in the recollected datasets, which is 80.0\%. This indicates that many of the tweets likely disappeared because of the removal of the user accounts, yet not having such a big impact in the elements contained in the tweet content, such as hashtags or URLs, as some of the textual content can be redundant across tweets.

\begin{figure}[t!]
 \centering
 \begin{subfigure}[t]{0.22\textwidth}
  \centering
  \includegraphics[height=1.5in]{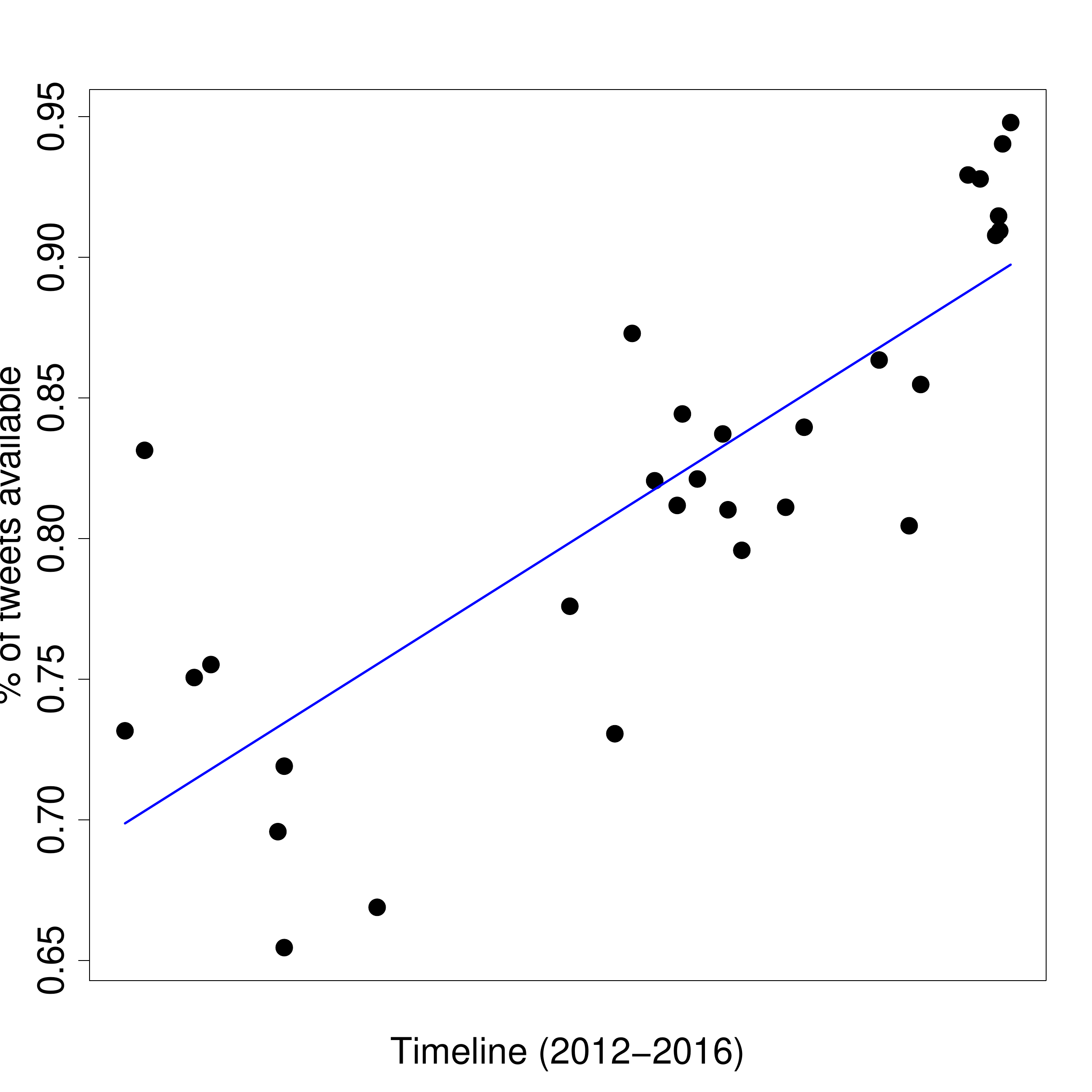}
  \caption{Completeness of tweets}
 \end{subfigure}
 ~
 \begin{subfigure}[t]{0.22\textwidth}
  \centering
  \includegraphics[height=1.5in]{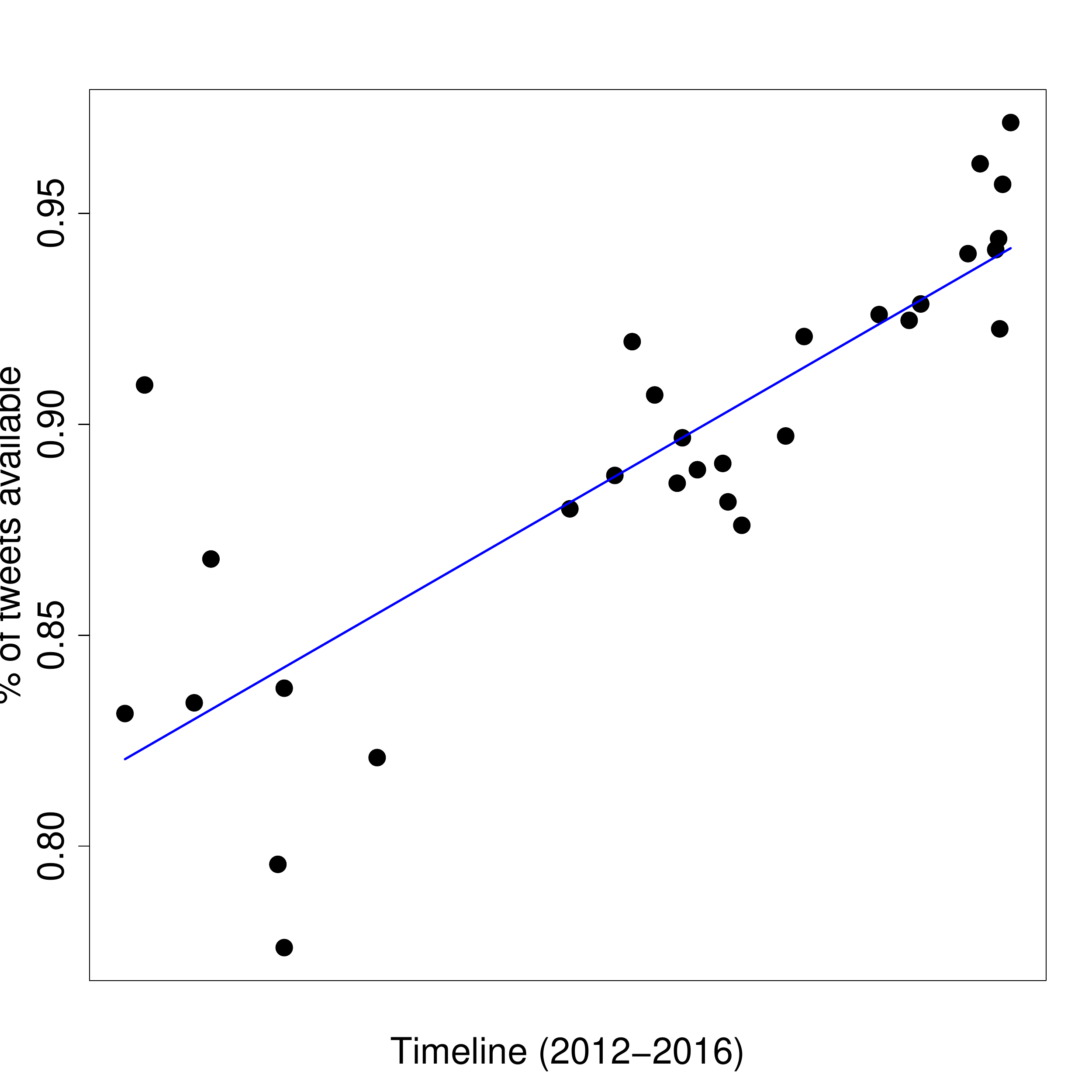}
  \caption{Completeness of hashtags}
 \end{subfigure}
 ~
 \begin{subfigure}[t]{0.22\textwidth}
  \centering
  \includegraphics[height=1.5in]{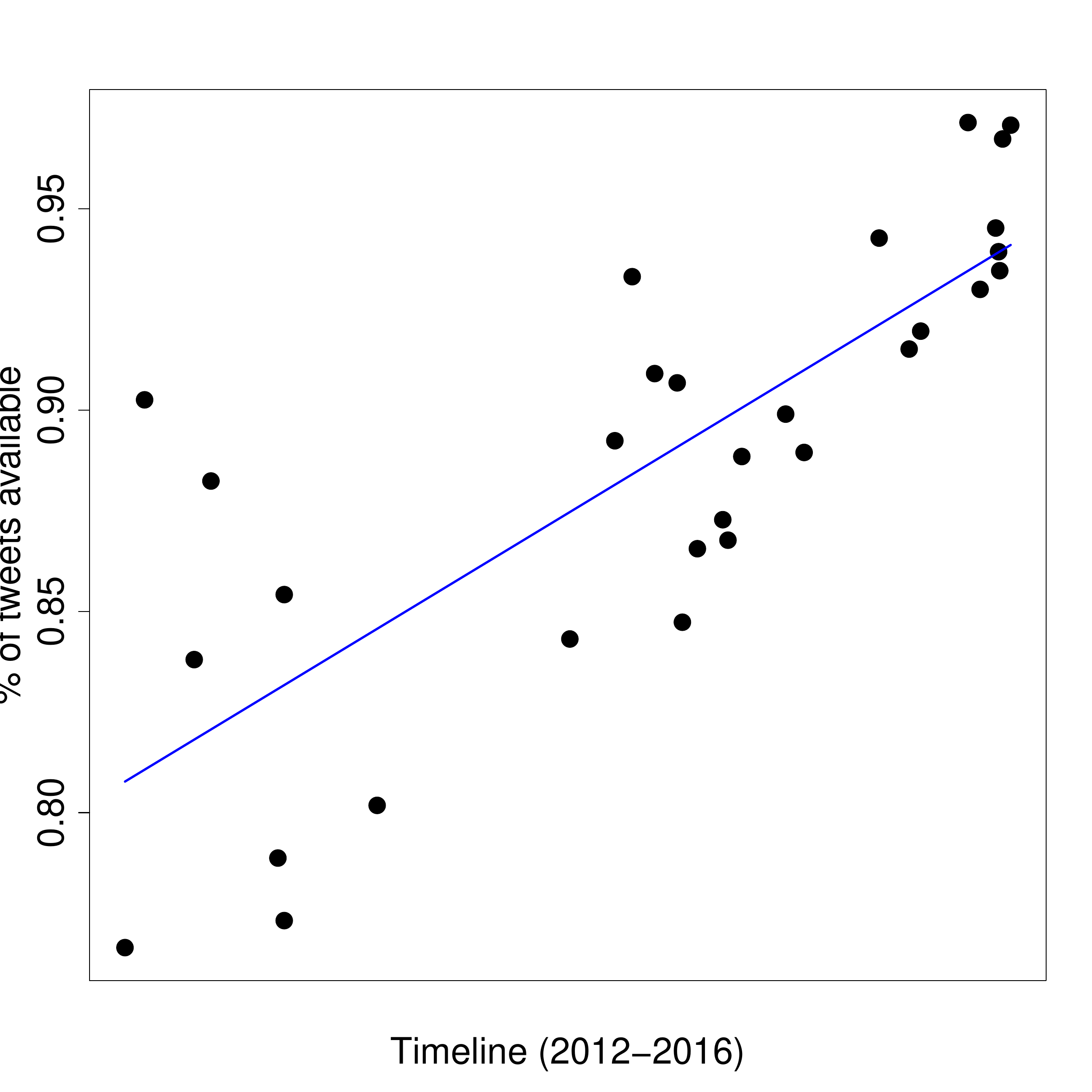}
  \caption{Completeness of mentions}
 \end{subfigure}
 \\
 \begin{subfigure}[t]{0.22\textwidth}
  \centering
  \includegraphics[height=1.5in]{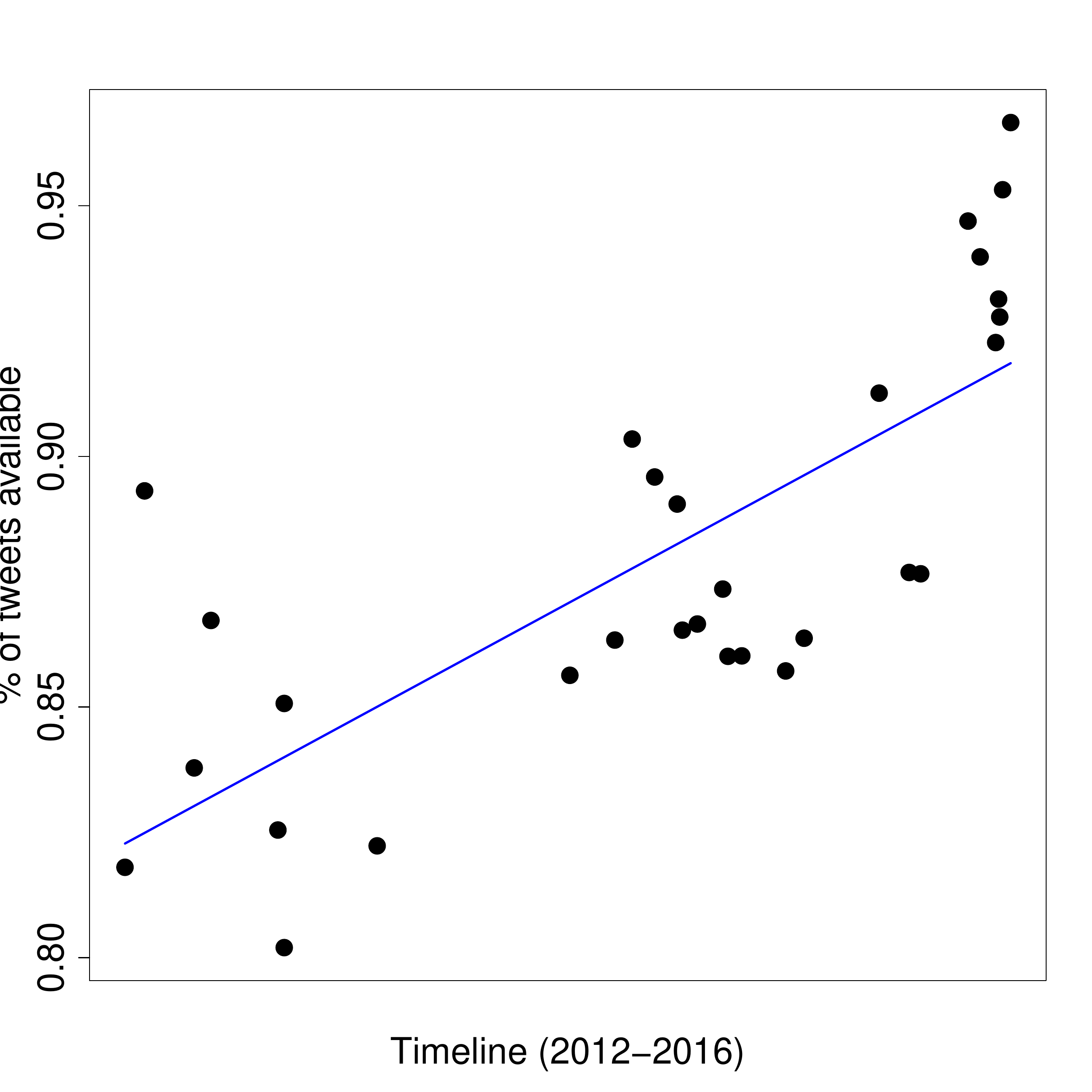}
  \caption{Completeness of tokens}
 \end{subfigure}
 ~
 \begin{subfigure}[t]{0.22\textwidth}
  \centering
  \includegraphics[height=1.5in]{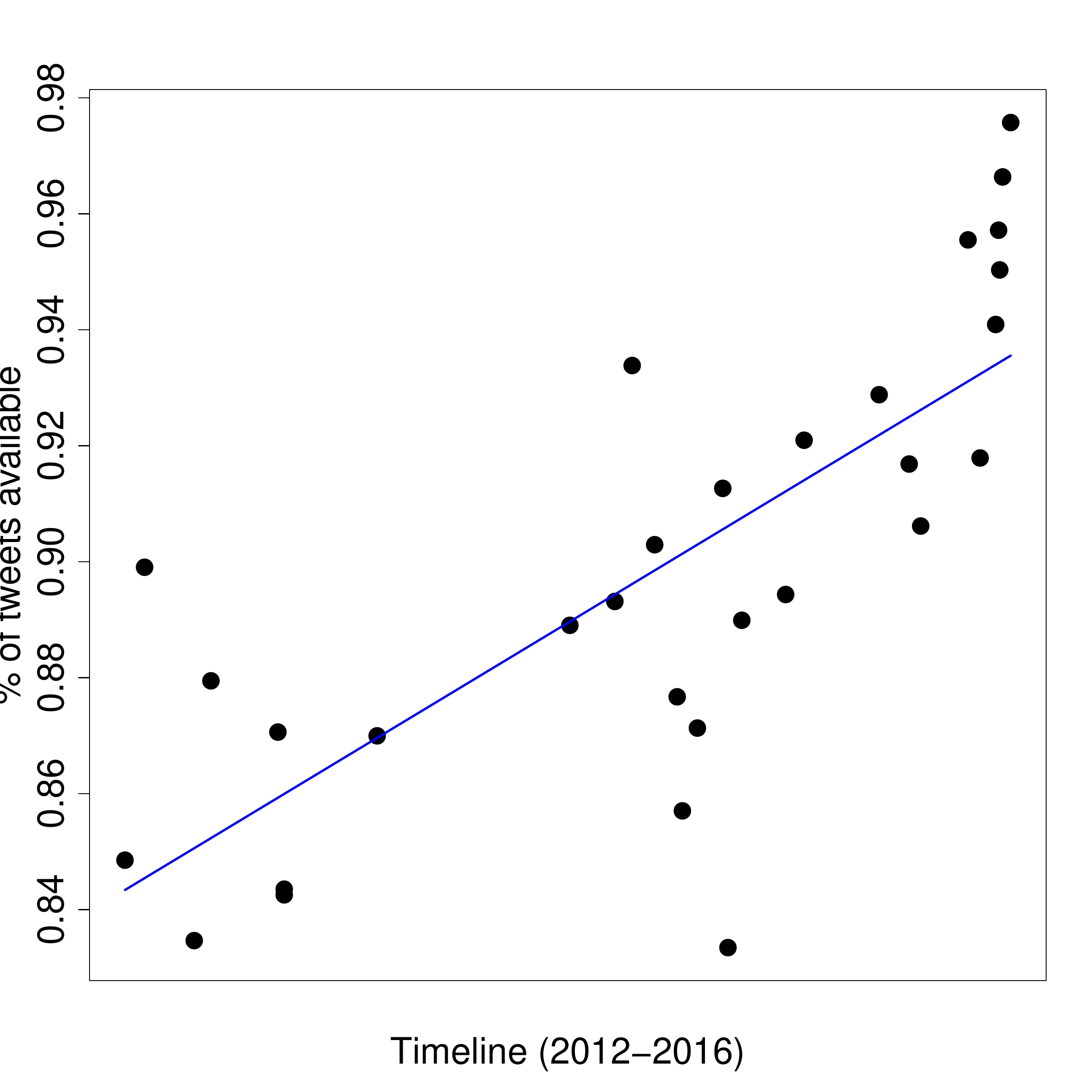}
  \caption{Completeness of URLs}
 \end{subfigure}
 ~
 \begin{subfigure}[t]{0.22\textwidth}
  \centering
  \includegraphics[height=1.5in]{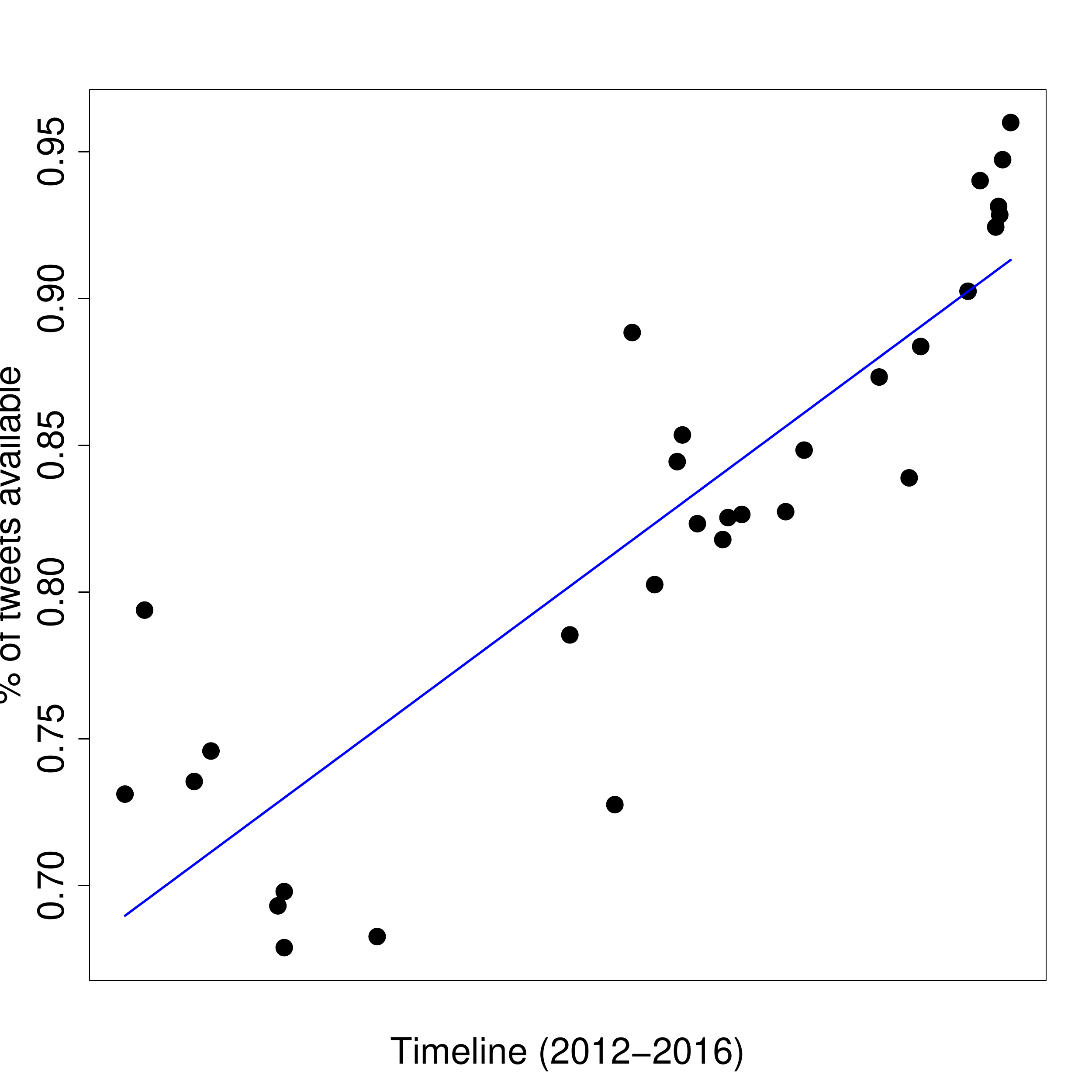}
  \caption{Completeness of users}
 \end{subfigure}
 
 \caption{Completeness by event, where events are sorted by time, with older events on the left and newer events on the right.}
 \label{fig:completeness}
\end{figure}

A detailed analysis of the available elements broken down by dataset is shown in Figure \ref{fig:completeness}. The figures show the degree of availability of different elements in each dataset, where the X axis represents the timeline (i.e. days elapsed from the first dataset). The blue line represents the line of best fit, showing the temporal tendency of the availability. The availability of all elements differs substantially when we look at different events, with older events having fewer elements available, as initially expected. They all show a similar trend, however with the completeness of tweets and users deteriorating to a greater extent, in some cases even dropping below 70\% of completeness. It is better for the rest of the elements, rarely falling below 80\%, showing again that the different elements that are part of the tweet content have higher degree of completeness thanks to the redundant content that can be found across different tweets. This leads to content disappearing from unavailable tweets being still accessible through other tweets with similar content.

\subsection{Representativity}
\label{ssec:representativity}

Despite the shortage of completeness, as we have observed, we want to measure if the data that is still available is representative of the original dataset. With the representativity we measure the extent to which the different features of a recollected dataset still represent those of the original dataset. For instance, we have lost some of the tweets, but is the remaining content still representative of the whole? Likewise, some users have disappeared, and the number of followers of the users that are still available may have changed; is the distribution of new follower counts still representative of the original distribution? Note that here we need to compare samples of different size, the original datasets being larger than the recollected ones.

To measure if recollected datasets represent the whole, we rely on Welch's t-test \cite{welch1947generalization} to measure if there is a statistically significant difference between two different populations, i.e. recollected features and original features. We do so for a range of 22 different features that are part of each tweet, and we group the datasets by year, hence having four different groups: 2012, 2013/2014, 2015 and 2016. We create vectors for each feature type as follows:

\begin{itemize}
 \item \textbf{Categorical features:} We create a vector where each value represents the number of occurrences of each categorical value. For instance, for languages (lang), we build a vector with the number of occurrences for each language, e.g. $v=\{|en|, |es|, |fr|, |de|\}$ (counts for English, Spanish, French, German)
 \item \textbf{Textual features:} We use a bag-of-words approach to create vectors representing each textual feature, i.e. each value of the vector representing the number of occurrences of a token.
 \item \textbf{Ordinal features:} Since Welch's t-test can only handle categorical values, we transform ordinal features into categorical by grouping its values by percentile, i.e. counting the number of occurrences in each percentile for a particular feature. For instance, if retweet counts range between 0 and 4,999 retweets, we split that range into 100 equal segments (0-49, 50-99, and so on) and count the number of occurrences in each segment, producing two vectors with 100 values each, one for the recollected features and one for the original features.
\end{itemize}

\begin{table}
 \centering
 \small
 \begin{tabular}{l l l l l}
  \hline
   & \textbf{2012} & \textbf{2013/2014} & \textbf{2015} & \textbf{2016} \\
  \hline
  text & 0.638 & 0.777 & 0.904 & 0.736 \\
  \hline
  hashtags & 0.975 & 0.933 & 0.877 & 0.982 \\
  \hline
  mentions & \textbf{0.000} & 0.085 & 0.723 & \textbf{0.016} \\
  \hline
  urls & \textbf{0.000} & \textbf{0.000} & \textbf{0.017} & \textbf{0.000} \\
  \hline
  authors & \textbf{0.000} & \textbf{0.000} & \textbf{0.000} & \textbf{0.000} \\
  \hline
  timestamps & \textbf{0.000} & \textbf{0.000} & \textbf{0.000} & \textbf{0.000} \\
  \hline
  favorite-count & \textbf{0.000} & \textbf{0.000} & \textbf{0.000} & \textbf{0.000} \\
  \hline
  lang & 0.846 & 0.946 & 0.985 & 0.996 \\
  \hline
  retweet-count & \textbf{0.000} & \textbf{0.000} & \textbf{0.000} & \textbf{0.000} \\
  \hline
  user-description & 0.197 & 0.110 & 0.281 & 0.837 \\
  \hline
  user-followers & \textbf{0.000} & \textbf{0.000} & \textbf{0.000} & \textbf{0.000} \\
  \hline
  user-following & \textbf{0.000} & \textbf{0.000} & \textbf{0.000} & \textbf{0.000} \\
  \hline
  user-lang & 0.986 & 0.997 & 1.000 & 0.996 \\
  \hline
  user-location & 0.825 & 0.415 & \textbf{0.000} & \textbf{0.000} \\
  \hline
  user-name & \textbf{0.000} & \textbf{0.000} & \textbf{0.000} & 0.617 \\
  \hline
  user-profile-image & \textbf{0.000} & \textbf{0.000} & \textbf{0.000} & \textbf{0.000} \\
  \hline
  user-screen-name & \textbf{0.000} & \textbf{0.000} & \textbf{0.000} & \textbf{0.000} \\
  \hline
  user-timezone & 0.525 & 0.691 & 0.831 & 0.967 \\
  \hline
  user-tweetcount & \textbf{0.000} & \textbf{0.000} & \textbf{0.000} & \textbf{0.000} \\
  \hline
  user-url & 0.948 & 0.953 & 0.966 & 0.992 \\
  \hline
  user-utc-offset & 0.625 & 0.684 & 0.836 & 0.967 \\
  \hline
  user-verified & 1.000 & 1.000 & 1.000 & 1.000 \\
  \hline
 \end{tabular}
 \caption{Representativity in p-values of recollected datasets with respect to original datasets, by year (p $<$ 0.05 are highlighted in bold).}
 \label{tab:representativity}
\end{table}

Table \ref{tab:representativity} shows p-values obtained through Welch's t-test for each pair of feature and dataset year. Values below 0.05 (highlighted in bold) indicate a statistically significant difference between the recollected and original values for a certain feature in datasets of a certain year. We observe that there is a big difference here across features, with some features showing statistically significant differences and hence lack of representativity; these include four types of features: (1) profile settings such as the user name, user profile image and screen name, which shows that users are likely to change those settings, (2) statistics associated with users, including tweet counts, number of followers and number of users they are following, which is expected to change as the users are active and keep tweeting and getting more followers, (3) statistics associated with tweets, where tweets' favourite and retweet counts change and are not representative of the original, and (4) tweet metadata, as the timestamps of the recollected tweets\footnote{Note that while timestamps of tweets do not change, here we are comparing the set of timestamps in the tweets available after recollection with the timestamps in the tweets in the original dataset.} as well as the tweet authors are not representative. On the other hand, there is a range of features which are representative of the original dataset, including user metadata such as the user's description, interface language, profile URL, UTC offset, or whether the account is verified are representative of the original dataset. Likewise, the content of the tweets is largely representative of the original, with the text and hashtags, some of the most used features for content-based research, being representative.

It is worth noting, however, that with the representativity we are measuring if the recollected features have the same distribution (i.e. average and standard deviation) as the original dataset. The lack of representativity does not imply that the recollected dataset does not show a similar trend as in the original data. For instance, the follower counts are expected to change and hence not be representative over time, however follower counts may keep growing proportionally for all users, leading to new follower counts that are not representative of the original, but still show a similar trend to that of the original. In the following section we look at the similarity of features to further explore this.

\subsection{Similarity}
\label{ssec:similarity}

With the similarity we aim to measure the extent to which different features of the recollected dataset have the same distribution as in the original dataset. It is important to note the difference between the representativity and the similarity in this analysis. While the representativity measures whether the recollected subset resembles the original dataset, with the similarity we are measuring if the recollected subset is similar to an equally sized subset of the original dataset. In the case of similarity, we are comparing two datasets of the same size. While for representativity the research question is ``does the available subset represent the whole?'', with the similarity we are interested in answering ``despite not having everything, does what we have resemble an equally sized, randomly sampled subset?''.

We need to create a large number of subsets of the original dataset, each equally sized as the recollected subset, to compare our recollected dataset with multiple random samples. Given the large size of the dataset, we use a stratified sample including a random 1\% of the tweets in each event. From this, we create 50,000 different random samples of the original dataset, and we compare each of them with the recollected dataset using the cosine similarity \cite{singhal2001modern}. The cosine similarity is an appropriate way of measuring the similarity between two equally sized populations, capturing whether the values of the features are similarly distributed (i.e. their vectors having a similar angle in the vector space model). We then average the cosine similarities obtained for all 50,000 random subsets. This enables us to compare the extent to which features in equally sized samples of the original and recollected datasets have similar distributions of values. The cosine similarity measures the extent to which two different sets of values are comparable, and it ranges from 0 (dissimilar) to 1 (similar). Again, as we do for the calculation of representativity, features representing ordinal values are converted into categorical vectors by computing the number of instances in each percentile, and textual features are represented using a bag-of-words approach.

Table \ref{tab:similarity} shows the cosine similarities for the 22 features, again grouped into 4 different years. This shows encouraging results compared to those obtained above for representativity. A look at the similarity values instead shows that most of the features are similar to the original dataset. This is the case with most of the features having similarity values above 0.9 consistently across all years. There are four exceptions, all of which are associated with the users' profile settings: the user profile image, the user screen name, the users' UTC offset and the user location.

Our results suggest that the recollected datasets are rather similar to randomly sampled subsets of the original datasets, however the distributions of values of some of the features are not representative of the whole.

\begin{table}
 \centering
 \small
 \begin{tabular}{l l l l l}
  \hline
   & 2012 & 2013/2014 & 2015 & 2016 \\
  \hline
  content & 0.999 $\pm 2.4e^{-5}$ & 1.000 $\pm 1.9e^{-5}$ & 1.000 $\pm 1.3e^{-5}$ & 1.000 $\pm 5.3e^{-6}$ \\
  \hline
  hashtags & 0.999 $\pm 3.1e^{-5}$ & 1.000 $\pm 9.8e^{-6}$ & 1.000 $\pm 1.6e^{-5}$ & 1.000 $\pm 2.5e^{-6}$ \\
  \hline
  mentions & 0.991 $\pm 3.5e^{-4}$ & 0.995 $\pm 2.0e^{-4}$ & 0.996 $\pm 1.6e^{-4}$ & 0.996 $\pm 1.8e^{-4}$ \\
  \hline
  urls & 0.952 $\pm 1.5e^{-3}$ & 0.975 $\pm 5.6e^{-4}$ & 0.954 $\pm 8.9e^{-4}$ & 0.988 $\pm 2.7e^{-4}$ \\
  \hline
  users & 0.759 $\pm 7.4e^{-4}$ & 0.861 $\pm 9.5e^{-4}$ & 0.856 $\pm 1.1e^{-3}$ & 0.952 $\pm 6.0e^{-4}$ \\
  \hline
  timestamps & 0.999 $\pm 3.1e^{-5}$ & 1.000 $\pm 8.9e^{-6}$ & 1.000 $\pm 1.3e^{-5}$ & 1.000 $\pm 1.7e^{-6}$ \\
  \hline
  favorite-count & 1.000 $\pm 5.6e^{-16}$ & 1.000 $\pm 6.5e^{-11}$ & 1.000 $\pm 7.7e^{-10}$ & 1.000 $\pm 5.3e^{-16}$ \\
  \hline
  lang & N/A & 0.996 $\pm 2.4e^{-5}$ & 1.000 $\pm 4.5e^{-6}$ & 1.000 $\pm 6.5e^{-7}$ \\
  \hline
  retweet-count & 1.000 $\pm 1.1e^{-7}$ & 1.000 $\pm 2.7e^{-6}$ & 1.000 $\pm 1.8e^{-6}$ & 1.000 $\pm 3.2e^{-7}$ \\
  \hline
  user-description & 0.980 $\pm 5.0e^{-4}$ & 0.994 $\pm 2.6e^{-4}$ & 0.977 $\pm 7.1e^{-5}$ & 0.957 $\pm 2.3e^{-4}$ \\
  \hline
  user-followers & 1.000 $\pm 1.9e^{-8}$ & 1.000 $\pm 2.0e^{-8}$ & 1.000 $\pm 5.0e^{-9}$ & 1.000 $\pm 2.8e^{-9}$ \\
  \hline
  user-following & 1.000 $\pm 2.0e^{-7}$ & 1.000 $\pm 5.3e^{-7}$ & 1.000 $\pm 6.0e^{-8}$ & 1.000 $\pm 1.3e^{-7}$ \\
  \hline
  user-lang & 1.000 $\pm 3.0e^{-6}$ & 1.000 $\pm 1.1e^{-6}$ & 1.000 $\pm 4.6e^{-6}$ & 1.000 $\pm 1.5e^{-6}$ \\
  \hline
  user-location & 0.967 $\pm 3.8e^{-4}$ & 0.989 $\pm 2.2e^{-4}$ & 0.510 $\pm 7.7e^{-4}$ & 0.215 $\pm 3.8e^{-4}$ \\
  \hline
  user-name & 0.943 $\pm 1.2e^{-3}$ & 0.960 $\pm 1.6e^{-3}$ & 0.960 $\pm 6.1e^{-4}$ & 0.996 $\pm 6.4e^{-5}$ \\
  \hline
  user-profile-image & N/A & 0.792 $\pm 8.6e^{-4}$ & 0.971 $\pm 9.9e^{-5}$ & 0.969 $\pm 1.3e^{-4}$ \\
  \hline
  user-screen-name & 0.588 $\pm 8.6e^{-4}$ & 0.777 $\pm 1.0e^{-3}$ & 0.794 $\pm 1.0e^{-3}$ & 0.921 $\pm 1.0e^{-3}$ \\
  \hline
  user-timezone & 0.971 $\pm 3.1e^{-4}$ & 0.990 $\pm 1.5e^{-4}$ & 0.997 $\pm 3.0e^{-5}$ & 1.000 $\pm 7.6e^{-6}$ \\
  \hline
  user-tweetcount & 0.992 $\pm 3.0e^{-5}$ & 0.995 $\pm 3.3e^{-5}$ & 0.999 $\pm 6.0e^{-5}$ & 1.000 $\pm 2.6e^{-6}$ \\
  \hline
  user-url & 1.000 $\pm 4.2e^{-8}$ & 1.000 $\pm 2.8e^{-8}$ & 1.000 $\pm 1.2e^{-8}$ & 1.000 $\pm 4.6e^{-8}$ \\
  \hline
  user-utc-offset & 0.803 $\pm 3.8e^{-4}$ & 0.981 $\pm 1.8e^{-4}$ & 0.910 $\pm 1.4e^{-4}$ & 0.986 $\pm 6.8e^{-5}$ \\
  \hline
  user-verified & 1.000 $\pm 1.1e^{-6}$ & 1.000 $\pm 1.1e^{-6}$ & 1.000 $\pm 3.3e^{-7}$ & 1.000 $\pm 1.4e^{-7}$ \\
  \hline
 \end{tabular}
 \caption{Similarity of recollected datasets with respect to equally sized random samples of the original datasets, by year. Reported values correspond to average cosine similarities and standard deviations, computed across 50,000 random samples. ``N/A'' indicates that the feature was not available in 2012 tweets as it was introduced later by Twitter.}
 \label{tab:similarity}
\end{table}

\subsection{Changingness}
\label{ssec:changingness}

To tackle the fourth and last research question, we look at the changingness of features, i.e. the extent to which the values of different features change from the original to the recollected dataset. One feature that never changes is the text of the tweet, as the text of a tweet remains the same as long as it is still available and can be recollected; it cannot be edited. Therefore we exclude the text and associated features such as hashtags or user mentions from this analysis. Likewise, the timestamp of a tweet does not change either, which is excluded from this analysis. In this analysis of changingness, we look at two different factors. On the one hand, we quantify the ratio of cases in which the value of a feature has changed. On the other hand, we further dig into those changes, which we analyse in three different ways depending on the type of feature:
\begin{itemize}
 \item \textbf{Ordinal features:} for ordinal features, we quantify the average increase of the feature values. For instance, for follower counts, we quantify, on average, how many new followers each user has gained.
 \item \textbf{Categorical features:} for categorical features, we list two values: the value that most frequently has changed from the original dataset, and the value that most frequently has been changed to in the recollected dataset.
 \item \textbf{Textual features:} in the case of textual features, we display the average Levenshtein distance \cite{levenshtein1966binary} of the changes, which measures the number of characters that have changed from the original value to the recollected value, on average.
\end{itemize}

\begin{table}
 \centering
 \small
 \begin{tabular}{l l l l l}
  \hline
  \multicolumn{5}{c}{\textbf{Ordinal features}} \\
  \hline
   & \textbf{2012} & \textbf{2013/2014} & \textbf{2015} & \textbf{2016} \\
  \hline
  favorite-count & 0.000 & 0.051 & 0.052 & 0.086 \\
   & (i: 0.000) & (i: 4.371) & (i: 6.517) & (i: 7.937) \\
  \hline
  user-tweetcount & 0.999 & 0.998 & 0.996 & 0.996 \\
   & (i: 20426.160) & (i: 20406.648) & (i: 17563.558) & (i: 3663.992) \\
  \hline
  retweet-count & 0.433 & 0.627 & 0.672 & 0.770 \\
   & (i: 361.398) & (i: 3933.477) & (i: 3066.591) & (i: 3901.538) \\
  \hline
  user-followers & 0.995 & 0.993 & 0.985 & 0.967 \\
   & (i: 7983.016) & (i: 3998.776) & (i: 3036.547) & (i: 422.558) \\
  \hline
  user-following & 0.985 & 0.981 & 0.955 & 0.916 \\
   & (i: 603.012) & (i: 445.826) & (i: 264.608) & (i: 81.623) \\
  \hline
  \multicolumn{5}{c}{\textbf{Categorical features}} \\
  \hline
   & \textbf{2012} & \textbf{2013/2014} & \textbf{2015} & \textbf{2016} \\
  \hline
  lang & N/A & 0.033 & 0.007 & 0.000 \\
   &  & (from: en) & (from: es) &  \\
   &  & (to: und) & (to: und) &  \\
  \hline
  user-lang & 0.021 & 0.015 & 0.016 & 0.004 \\
   & (from: en) & (from: en) & (from: en) & (from: en) \\
   & (to: en) & (to: en) & (to: en) & (to: en) \\
  \hline
  user-profile-image & N/A & 0.717 & 0.542 & 0.250 \\
   &  & (from: custom) & (from: custom) & (from: custom( \\
   &  & (to: custom) & (to: custom) & (to: custom) \\
  \hline
  user-timezone & 0.133 & 0.085 & 0.054 & 0.012 \\
   & (from: None) & (from: None) & (from: None) & (from: None) \\
   & (to: Eastern Time) & (to: Pacific Time) & (to: Pacific Time) & (to: Pacific Time) \\
  \hline
  user-utc-offset & 0.663 & 0.192 & 0.426 & 0.128 \\
   & (from: -18000) & (from: None) & (from: 3600) & (from: 0) \\
   & (to: -14400) & (to: -14400) & (to: 7200) & (to: 3600) \\
  \hline
  user-verified & 0.016 & 0.008 & 0.004 & 0.001 \\
   & (from: False) & (from: False) & (from: False) & (from: False) \\
   & (to: True) & (to: True) & (to: True) & (to: True) \\
  \hline
  \multicolumn{5}{c}{\textbf{Textual features}} \\
  \hline
   & \textbf{2012} & \textbf{2013/2014} & \textbf{2015} & \textbf{2016} \\
  \hline
  user-description & 0.748 & 0.681 & 0.594 & 0.415 \\
   & (l: 79.342) & (l: 65.194) & (l: 51.097) & (l: 29.150) \\
  \hline
  user-location & 0.394 & 0.253 & 0.240 & 0.372 \\
   & (l: 14.110) & (l: 14.226) & (l: 10.862) & (l: 5.573) \\
  \hline
  user-name & 0.447 & 0.345 & 0.253 & 0.081 \\
   & (l: 10.383) & (l: 10.504) & (l: 10.356) & (l: 10.430) \\
  \hline
  user-screen-name & 0.222 & 0.128 & 0.077 & 0.016 \\
   & (l: 8.952) & (l: 9.207) & (l: 9.034) & (l: 9.044) \\
  \hline
  user-url & 0.492 & 0.384 & 0.324 & 0.328 \\
   & (l: 23.840) & (l: 18.581) & (l: 21.346) & (l: 23.259) \\
  \hline
 \end{tabular}
 \caption{Changingness of tweet metadata for each pair of feature and dataset year. The table also shows the average increase (i) of the feature value for ordinal features, the most frequent disappearing (from) and appearing (to) features for categorical features, and the average Levenshtein distance (l) of the change for textual features.}
 \label{tab:changingness}
\end{table}

Table \ref{tab:changingness} shows the results of the analysis of the changingness of features. We can observe that ordinal features are very likely to change, especially users' tweet counts, number of followers and followees, which is expected for user accounts that are active. They actually vary substantially irrespective of the dataset being older or newer. The favourite and retweet counts of tweets also keep varying, although it is less common for the number of favourites, with fewer than 10\% of the tweets changing.

Among the categorical features, we observe that most of them are affected as the dataset gets older. Features like a user's profile image or a user's UTC offset are very likely to change for old datasets in more than 60\% of the cases. Other features, such as a user's interface language and whether users are verified, are very unlikely to change.

In the case of the textual features, all of which are related to users' profile settings, we observe that they are increasingly likely to be modified as the datasets get older. The users' screen names (i.e. the Twitter handles) are the least likely to change, still more than 22\% of the users changed it since 2012. The rest of the features, which include the users' website/URL, the users' real name, their location and description, are very likely to change, with variability values between 39\% and 75\%.


\section{Discussion}
\label{sec:discussion}

We have looked at 30 Twitter large-scale, longitudinal datasets originally collected over the course of four years (2012--2016), and compared them with datasets recollected in 2016. This analysis is important as researchers aiming to recollect datasets shared by others through tweet IDs end up collecting a reduced subset of the original dataset, whose metadata may have changed. Our longitudinal study of the completeness, representativity, similarity and changingness of recollected datasets has enabled us to quantify the extent to which recollected datasets resemble the original ones.

To tackle this analysis, we set forth four research questions, as follows:

\begin{description}
 \item[RQ1] \textit{How complete is a dataset recollected from lists of publicly available tweet IDs? (completeness)} \\
 Our analysis buttresses the expected tendency that tweets keep disappearing over time, with older datasets having fewer tweets available than newer ones. The ratio of tweets and unique users available in the recollected dataset can drop below 70\% within four years. However, this is less dramatic for other elements found in tweets, such as hashtags or URLs, whose availability is around 80\% after the same time period. The redundancy of textual content across tweets makes it possible for more unique hashtags and URLs to be available after a long period of time.
 \item[RQ2] \textit{How representative is the recollected dataset? (representativity)} \\
 The textual content of the recollected datasets has been found to be largely representative of the original datasets, despite being a reduced subset. We have found that the difference between the textual content of the recollected subsets and that of the original datasets is not statistically significant. However, we found statistically significant differences for most of the tweets' metadata, showing that the recollected metadata is no longer representative of the original.
 \item[RQ3] \textit{How similar is the recollected subset to a randomly sampled subset? (similarity)} \\
 Given that the recollected datasets are smaller, we have created 100 random subsets of the original datasets, whose size matches that of the recollected datasets. With these 100 random samples, we have computed the similarity of the different features with respect to the recollected dataset. In this case, we have found substantial similarities for the majority of the features, which shows that despite being reduced subsets of the whole, they are still largely similar to a random subset of the original data.
 \item[RQ4] \textit{How has the metadata associated with the successfully recollected subset changed? (changingness)} \\
 We have looked at the ratio of cases in which the different metadata conforming the tweets have changed. We have observed big differences in the case of the most of the features, including the majority of features associated with the users' profile settings. There are only a few exceptions of features that are less likely to vary over time, which include the users' interface language, whether the users are verified and the tweets' favorite counts. The features that are most likely to change over time include the users' followeer and followee counts, the users' tweet counts and the users' descriptions.
\end{description}

\textbf{Summary and Implications of the Study.} Our analysis shows that datasets gradually keep shrinking as tweets become unavailable over time, showing that up to 30\% of the tweets can disappear within four years. This is however not as dramatic when it comes to the diversity of elements that we observe, such as the set of unique URLs or hashtags, thanks to those elements being redundant across tweets and being less affected by the deletion of a subset of the tweets. Our analysis comparing the characteristics of original and recollected datasets shows encouraging conclusions especially when it comes to the textual content of tweets, as the textual content that remains available over time is still largely representative of and similar to the original dataset. This is positive for all the research that relies mainly on the tweets' textual content. Results are not as positive when we look at other tweet and user metadata. Our analysis suggests that most metadata are similar to that of a randomly sampled subset of the original dataset, however metadata tend to change and they are no longer representative of the original dataset. It should therefore be expected that the metadata of the recollected dataset resembles that of a randomly sampled subset, but not necessarily the original dataset. This should be taken into account as it may have a significant impact on the outcome of a research study, depending on the objectives and how the data is looked at. Studies looking at the metadata of recollected Twitter datasets should take into consideration the effect that the change of metadata may have had in their analysis.

Our analysis shows that researchers relying on datasets recollected from tweet IDs can largely assume that the dataset is similar to a randomly sampled, equally sized subset of the original dataset. Since the researcher has the list of tweet IDs, they can also know the percentage of tweets that has been successfully collected. When a researcher manages to recollect a $p$\% of the original dataset, they can then expect to have a dataset that is similar to a random sample containing $p$\% of the original tweets. The recollected dataset, however, cannot be deemed representative of the whole, original dataset. Where a researcher wants to analyse metadata of the tweets, they should be particularly careful if they want to analyse ordinal features (e.g., followers, followees, user tweet counts, retweet counts), as these features present the highest probability to change over time. Features associated with user preferences (e.g. user language, timezone) are less likely to change. Other textual features in user settings (e.g. user description, user name, user location) are also likely to change, challenging their consistency over time.

While this study shows that researchers can use recollected datasets for their research if the above limitations are carefully considered, researchers should always report the source of the tweet IDs used in their research, as well as the percentage of tweets that they successfully managed to collect, which is an important metric to quantify the resulting dataset.

This study shows the extent to which researchers can rely on newly recollected datasets from publicly shared tweet IDs. It has focused particularly on Twitter datasets associated with real world events. An aspect that has not been covered in this study is the replicability of datasets built from user profiles, e.g. including timelines of tweets for a set of users. A longitudinal study of the persistence of Twitter datasets built from user timelines would be an interesting complementary analysis that we leave for future work.


\bibliographystyle{ACM-Reference-Format}
\bibliography{tweet-recollection}   

\end{document}